# Feasibility of cmOCT angiographic technique using 200 kHz VCSEL source for *in vivo* microcirculation imaging applications

CERINE LAL[1], HREBESH M SUBHASH[2], SERGEY ALEXANDROV[1], MARTIN J LEAHY *[1,3]

[1] *Tissue optics and microcirculation imaging Facility, National Biophotonics and Imaging Platform, National University of Ireland, Galway*
[2] *Colgate-Palmolive Global Technology Centre, 909 River Rd, Piscataway, NJ 08854, USA*
[3] *Royal College of Surgeons (RCSI), Dublin, Ireland*
*Corresponding author: martin.leahy@nuigalway.ie

**Abstract**

Optical coherence tomography (OCT) angiography is a well-established *in vivo* imaging technique to assess the overall vascular morphology of tissues and is an emerging field of research for the assessment of blood flow dynamics and functional parameters such as oxygen saturation. In this study, we present a modified scanning based correlation mapping OCT (cmOCT) using a 200 kHz high speed swept source OCT system operating at 1300 nm and demonstrate its wide field imaging capability in ocular angiographic studies.

## 1. INTRODUCTION

Microcirculation refers to blood flowing through the circulatory system of an organism comprising of network of vessels (arterioles, capillaries and venules) that are typically less than 100 μm in diameter [1]. Changes in microcirculation can reflect an underlying change in tissue metabolism or the response of the tissue to a given treatment or an exposure to a stimulant. Over the years, many techniques have been demonstrated for measuring blood flow and related microvascular parameters [1], [2]. Among these, light based techniques have garnered research interests due to its noninvasive and high-resolution imaging capability. Of these, OCT is a promising imaging modality for microcirculation imaging that can reconstruct three dimensional images of vascular structures with sufficient axial and lateral resolution to a depth of 3 mm in vivo in the tissues. One of the major advantages of OCT is that along with structural information of the biological sample of interest, it can also be used to extract functional information from the same data without disturbing the sample [3]. Imaging the microcirculation helps scientists and clinicians to understand the microvasculature, both structural and functional, whereby early detection in its abnormality can lead to early diagnosis and treatment efficacy. Among the functional variants, angiographic OCT enables visualization of blood vessels from the acquired OCT B- scan images and has been of high research interest in the recent past with more than 1000 publications per year over the past five years. OCT angiography refers to producing microvasculature maps from the acquired OCT interference spectrum [4]. As OCT detects ballistic photons that are backscattered from the various regions within the tissue, any change in phase or amplitude of the backscattered photons are reflected in the interference spectra. OCT angiographic techniques detect these shifts in phase or amplitude of the backscattered light from the moving red blood cells (RBC) within the blood vessels over time compared to the static tissue background. Based on the detection techniques, OCT angiographic techniques can be divided into phase-based, amplitude-based and techniques that use both amplitude and phase of the interference signal [5]. These include speckle variance [6], correlation mapping [7], optical microangiography (OMAG) [8], phase variance [9], and Doppler OCT [10, 11]. The amplitude -based techniques work on the principle of detecting changes in the OCT intensity images (B scans) over time to generate contrast between the static and dynamic structures within the tissue [12]. Similarly, the phase - based techniques detect the changes in the phase or the extent of decorrelation between the phase of the Fourier transformed spectral data acquired at the same location over time to achieve motion contrast [13]. Phase based angiographic techniques are further divided into Doppler angiography [10], phase variance [9] and techniques that make use of both intensity and phase information. These complex OCT angiographic techniques include optical microangiography [8], complex differential variance [14], and methods based on complex correlation of the interference spectra [15, 16]. Though amplitude and phase-based techniques provide motion contrast and are sensitive to the movement of RBC's, phase - based techniques are highly sensitive to system noise and axial motion in the sub - wavelength region and require complex phase correction methods to eliminate the artefacts [17-19]. Amplitude-based OCT techniques are less sensitive to bulk tissue motion, however they are dependent on the back scattered intensity from different regions within the sample which may deteriorate the motion contrast in highly scattering tissue. Slower image acquisition speed of OCT systems was one of the major limiting factors for the suitability of OCT angiography in clinical studies [20]. Slower speeds introduced motion artefact during image acquisition caused by the bulk motion of the sample. With the introduction of Fourier domain OCT (FD-OCT) systems especially swept source OCT (SS-OCT), imaging speed in the range of MHz has been achieved [18], [21]–[27]. Also compared to super luminescent diodes (SLD's) which were used in spectral domain OCT systems, swept source lasers offer higher scanning speeds, extended depth range with reduced sensitivity roll-off, reduced fringe washout and improved light detection efficiency due to dual balanced detection [28]. SS-OCT's have been developed based on either VCSEL (vertical-cavity surface-emitting laser) lasers, FDML (Fourier domain mode locked) laser sources or akinetic lasers. Apart from a recent study that used phase stabilised akinetic laser source [22] for phase difference angiographic imaging, other phase - based algorithms based on swept sources required complex post processing. Hence most of the swept source based angiographic techniques are intensity based. Recent studies involving phase gradient based phase stabilisation techniques have been proposed for SS – OCT systems [29, 30] which does not require the correction of bulk motion and jitter induced phase artifacts for complex field based angiography. However, they still require pre-processing steps compared to intensity based techniques and also the accuracy of phase stabilization depends on the signal intensity of each acquired A-line. Among the intensity based OCT angiography, techniques such as speckle variance [6], correlation mapping OCT [7] and split spectrum amplitude decorrelation (SSADA) [31] have been successfully implemented using SS-OCT. SV was first demonstrated in 2008 using FDML based SS-OCT operating at 36 kHz A-scan rate [6]. It was shown that an optimal signal to noise ratio was achieved using a time window of 8 B-scans at each transverse position [32]. SSADA technique introduced in 2012 [31], reported improved sensitivity to flow detection by reducing the axial resolution of the OCT voxel compared to its transverse resolution for retinal imaging applications where retinal and choroidal blood flow is mostly in the direction perpendicular to the sample beam. It was introduced as an enhancement of the amplitude- decorrelation method so that axial decorrelation sensitivity is reduced for retinal angiographic applications. The technique used 8 B-scans to calculate the decorrelation between subsequent frames at each transverse location obtained from a short cavity swept laser based SS-OCT system operating at 100 kHz A- scan rate. It requires more pre processing steps where the full k space spectrum to be divided into different sub bands, depending on the ratio of axial to spatial resolution. cmOCT mapping introduced in 2011 [7], also uses time varying properties of speckle statistics to differentiate between the static and dynamic regions. However, cmOCT is a kernel based approach that finds the extent of decorrelation between successive B-scans using Pearson's correlation coefficient or similar. The extent of decorrelation between the adjacent B-scans is determined by the flow through the region. The first studies based on cmOCT were demonstrated using VCSEL based SS-OCT systems operating at an A-scan rate of 16 kHz [33-35]. Spectral interferograms were acquired using a dense scanning protocol that ensured that the inter-frame separation was within the transverse resolution of the OCT system to insure strong correlation between adjacent B-scans. This led to longer data acquisition time restricted its use to small field of imaging applications compared to other swept source based intensity angiographic techniques.

In this study, we investigate microvascular detection capability of cmOCT at high scanning speed using 200 kHz VCSEL based SS-OCT system for *in vivo* imaging applications. To test the performance of cmOCT algorithm, first it is tested on intralipid phantom model and then is extended to nailfold imaging and to preclinical small animal imaging applications.

## 2. EXPERIMENTAL SETUP AND METHODS

*2.1 High speed SS-OCT system*

For conducting the experiments, commercial SS-OCT system from Thorlabs operating at 200 kHz was used. It uses micro-electro-mechanical systems vertical-cavity surface-emitting (MEMS-VCSEL) swept laser operating at central wavelength of 1300 nm (spectral bandwidth of 100 nm) and has an imaging depth range of 12 mm. The system uses a 5X objective that provides a spatial resolution of 25μm. The measured system sensitivity was 100 dB near the zero delay line. The system was capable of imaging using multiple B-scan (B-M) protocol which let the user acquire multiple B-scans at the same lateral position.

*2.2 cmOCT processing and data acquisition*

Details of cmOCT processing algorithm are as explained in previous papers. Briefly, cmOCT algorithm is a two-step process that uses the magnitude of the Fourier transformed spectral signals to reconstruct the microvascular map from the static tissue background. The cmOCT algorithm is given by Eq.1.

$$cmOCT(x,y) = \sum_{p=0}^{M}\sum_{q=0}^{N} \frac{[I_A(x+p,y+q)-\overline{I_A(x,y)}][I_B(x+p,y+q)-\overline{I_B(x,y)}]}{\sqrt{[(I_A(x+p,y+q)-\overline{I_A(x,y)})]^2 + [(I_B(x+p,y+q)-\overline{I_B(x,y)})]^2}}$$

(1)

where $I_A$ and $I_B$ are the OCT intensity images at 2 successive images at same location and $M$ and $N$ are the kernel size. $\overline{I_A(x,y)}$ and $\overline{I_B(x,y)}$ denote the mean intensity within the $M$ x $N$ kernel size and *x, y* denote the pixel location within the kernel of interest.

cmOCT is a kernel based approach where the kernel of chosen size is shifted across the entire image to produce the corresponding 2D correlation map between adjacent B-scans. The resulting correlation map contains values in the range of 0 ± 1 indicating weak or strong correlation with zero denoting highly decorrelated data between the pair of images considered. Static regions of the tissue return a high correlation value (> ~0.6) whereas vascular areas containing blood vessels with flowing blood returns low correlation value (~ - 0.6). As we are interested in the dynamic regions where there is blood flow, for visualization purpose, the decorrelation map is inverted with static regions represented by low intensity (dark) and regions of flow represented by higher intensity values (bright). In the next step to supress background noise from the static regions, a binary structural mask is applied based on a set threshold onto the generated 2D correlation map. Finally, to enhance the regions of flow and suppress the static regions, the resulting cmOCT values are mapped onto the range -0.6 to +0.6 using equation (2)

If we have numbers *x* in the range *[a, b]* and we want to transform them to numbers *y* in the range *[c, d]*, then

$$y = (x-a)\frac{d-c}{b-a} + c \qquad (2)$$

For the present paper, we acquired 8 B-scans at each location to produce cmOCT map following the multiple B scan (B-M) acquisition protocol following the swept source based speckle variance and SSADA techniques [32], [31]. cmOCT algorithm was applied to the successive B-frames within these 8 frames thus resulting in *N*-1, 2 D decorrelation maps (*N*=8 in this case) at a given location. Finally, the correlation maps are averaged to produce averaged cmOCT maps as given by Eq. (3).

$$Averaged_{cmOCT}(x,y) = \frac{1}{K}\sum_{K=1}^{7} cmOCT(x,y) \qquad (3)$$

The cmOCT map was generated using a custom written software based on Java [35]. A kernel size of 5 × 5 was empirically chosen to obtain optimal performance with the trade-off of processing time and motion contrast as shown in table 1.

To compare our proposed cmOCT algorithm for 200 kHz SS-OCT to speckle variance imaging, SV algorithm was implemented in MATLAB.

The inter-frame speckle variance (*SV*) based on B-M scan protocol is given by Eq. (4)

$$SV(x,z) = \frac{1}{N}\sum_{i=1}^{N}(I_i(x,z) - I_{mean})^2 \quad (4)$$

where $I_i(x,z)$ is the $i^{th}$ B-scan with lateral and depth indices *x* and *z*. $I_{mean}$ is the mean intensity of a given pixel over the *N* scans (*N* = 8 in this case).

### 2.3 Post processing for motion artefact removal

It is highly desirable to minimize the motion artefact arising from the involuntary sample movement or the inherent rhythmic vibrations within the organism for in vivo imaging of small animal models and clinical applications [19]. These movement artefacts introduce error in the cmOCT maps thereby reducing the image contrast between static and flow regions and thus degrade the signal to noise ratio. Although we used 200 kHz axial scanning system for data acquisition, bulk motion artefacts may occur and hence we have used axial and lateral motion correction algorithms before applying correlation mapping. To compensate the axial motion between the sequential B scans acquired from the same location, we implemented the algorithm described by Makita et al [36] which is based on one dimensional cross correlation between the intensity profiles of A-lines within the successive B frames acquired from the same location. After axial motion compensation, the image registration method developed by Thevenaz et al [37] was implemented to compensate for the lateral motion occurring between the B-scans,. Finally, after producing cmOCT maps, median filtering of the resulting correlation maps using a 3x3 kernel was carried out to improve the signal to noise ratio of the resulting images. The same pre-processing steps that were carried out for cmOCT was done for SV angiography. Figure 1 shows the schematic of the processing steps involved for both cmOCT and SV processing.

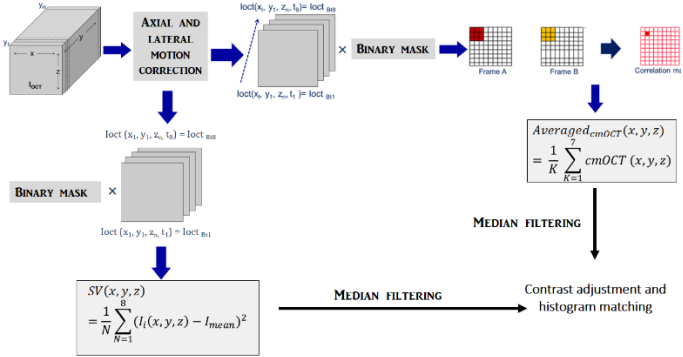

Fig.1. The schematic showing cmOCT and SV processing steps

### 2.4 Measure of contrast

In order to compare between cmOCT and SV results on the obtained angiographic maps, we have used metrics correlation signal to noise ratio (CSNR) [38, 39], signal to noise ratio measurement as described in [31, 40] and contrast to noise ratio (CNR) [29].
CSNR is defined as in Equation (5) [38]

$$CSNR = \frac{\overline{C_{FLOW}} - \overline{C_{STATIC}}}{\sqrt{\sigma^2_{STATIC}}} \quad (5)$$

Where $\overline{C_{FLOW}}$ and $\overline{C_{STATIC}}$ are the average correlation values in flow and static regions respectively, and is $\sigma^2_{STATIC}$ is the variance of the correlation within the static region.

To calculate the SNR as given in [38, 39], the standard deviation of signals within the static region was calculated, and was used as a threshold for defining effective detectable flow signals within the flow region. CNR is measure using Equation (6) given by [29]

$$CNR = \frac{\overline{C_{FLOW}} - \overline{C_{STATIC}}}{C_{STATIC}} \quad (6)$$

To compare between cmOCT and SV images using profile plots, we have scaled both cmOCT and SV values to the range 0 to 1 using Eq. (2).

We have provided depth color coded projection images for both enface cmOCT and SV processed angiographic images. The depth encoded colour intensity images provide visualisation of blood vessels within the overlapping regions in depth which is not visible from maximum intensity projection images. The depth encoded intensity images were generated in Matlab (version 2014). For the depth encoding, each image slice was colour coded using the command 'hsv2rgb' by providing the set hue, saturation and value. For each image in the volume stack, the value of hue was incremented by fixed step size determined by the total number of frames in the volume stack. In this way, we would have mixed colours on overlapped vessel areas which will be missed if we used maximum intensity projection based depth encoding. However, the depth encoding provides only visual information and is not used for comparison between cmOCT and SV techniques.

## 3   RESULTS AND DISCUSSIONS

To demonstrate the feasibility of cmOCT for high speed SS-OCT, we first performed experiments using capillary tubes filled with 2 % of 20 % v/v intralipid solution embedded in synthetic clay to mimic the Brownian motion of dynamic scattering particles. The capillary tube with an inner diameter of ~300µm was embedded in Blu-Tack to simulate the static background tissue optical heterogeneity. Three dimensional (3D) volumetric data covering an area of 3 x 5 mm² with 8 B-scans at each location was acquired. In the Y scan direction, 200 scan locations covering 5 mm were acquired with 1000 A-lines in each B-scan, thus generating a total volume of 1600 images using the B-M scan protocol. The total time for 3 D volume acquisition was 27 seconds. The spectral interferograms were processed to produce intensity B-scan images following conventional Fourier domain processing. The processed intensity images were subjected to axial and lateral motion compensation as discussed in section 2.3 and finally processed using averaged cmOCT angiographic technique. As mentioned in section 2.3, a kernel size of 5×5 was chosen for the cmOCT processing for *in vivo* imaging applications based on the obtained contrast measurement parameters as indicated in table 1 for capillary flow phantom experiment.

Table 1 Comparison metric for motion contrast for phantom experiment for different kernel sizes

| Comparison Metric | Kernel size (3 ×3) | Kernel size (5 ×5) | Kernel size (7 ×7) |
|---|---|---|---|
| CSNR | 5.1 | 11.1 | 12.5 |
| SNR | 77.1 dB | 91.8 dB | 87.8 dB |
| CNR | 394.4 | 3.1055 × 10³ | 1.470 × 10³ |

Table 2. Comparison metric for motion contrast for phantom experiment

| Comparison Metric | cmOCT | SV |
|---|---|---|
| CSNR | 11.1 | 4.7 |
| SNR | 91.8 dB | 73.2 dB |
| CNR | 3.1055 × 10³ | 232.3 |

Figure 2 shows the result of modified scanning protocol based cmOCT algorithm on the capillary phantom model using a kernel size of 5×5. Figure 2 (a) shows the representative B-scan image of the intralipid filled capillary tube embedded in Blu Tack and figure 2(b) and 2(c) shows maximum intensity projection (MIP) of enface images obtained using cmOCT and SV technique respectively. Figure 3(a) and 3(b) shows the cmOCT and SV maps obtained from the B scan image shown in figure 2(a). From figure 3(a), it can be observed that cmOCT effectively supresses the background noise from static regions and produce bright pixel values in regions where dynamic scattering over time occurs. Figures 3(c) and 3(d) shows the flow profiles extracted from the regions marked (white line) in 3(a) and 3(b) respectively, with SV and cmOCT values mapped onto the same scale range (0 to 1) for comparison. The results obtained from the motion contrast analysis is shown in table 2. From table 2, we can observe that there is an improvement of approximately two-fold in CSNR and 20 dB in SNR calculation for cmOCT compared to SV technique with the current scanning protocol.

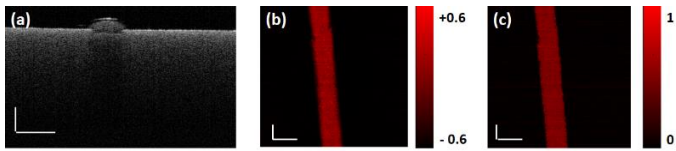

Fig. 2. (a) B scan image of intralipid filled capillary tube embedded in synthetic clay, (b) MIP of enface cmOCT images (c) MIP of enface SV images (Scalebars in x and y direction is 500 μm).

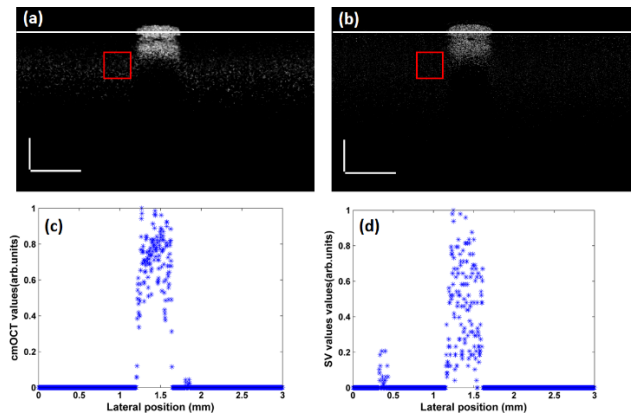

Fig. 3. (a) cmOCT image obtained from the transverse scan shown in Fig. 2(a) with static region of interest marked in red, (b) corresponding SV image with static region of interest marked in red (c) Profile plot obtained over the marked line in (a), (d) Profile plot obtained over the marked line in (b) (scale bar is 500 μm).

From table 2, we observe that cmOCT has better motion contrast compared to SV in flow phantom experiments.

To test the performance of cmOCT for high speed angiographic imaging in high motion *in vivo* scenario, we imaged the vasculature of nailfold of a healthy human volunteer. For the study, the little finger of the volunteer was fixed on a home built imaging mount so that the dorsal skin surface faced the OCT probe beam. Imaging was performed using the B-M scan protocol over an area of 5 mm × 3 mm. The scanning protocol was set to capture 1000 A-scans per B-scan over a scan distance of 5 mm. For 3D volumetric imaging, we acquired 300 B-scan sampling positions over the set scan range of 3 mm and 8 repeated B-scans were captured at every C-scan position thus generating 2400 B scans per volume. Figures 4 (a)-(e) shows the results obtained from the nailfold imaging studies. Figure 4 (a) is the representative structural image of the nailfold and 4 (b) and 4 (c) shows the corresponding cmOCT and SV maps. Figures 4(d) and 4(e) shows the depth encoded intensity projection image obtained from enface cmOCT and SV images respectively. From figure 4 (d), it can be observed that the cm-OCT technique clearly maps the hair pin shaped capillary loops within the nail-fold using the present high-speed system. However, the lateral resolution of the system limits the resolvability of individual capillary loops.

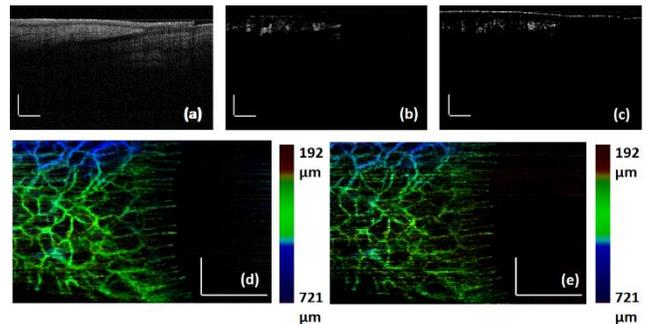

Fig. 4. (a) Representative B-scan image of the little finger of adult volunteer, (b) corresponding cmOCT image, (c) corresponding SV image (d) depth encoded intensity projection image of enface cmOCT images, (e) depth encoded intensity projection image of enface SV images (scale bars in x and y directions are 500 μm ).

Comparing the MIP's of cmOCT and SV in figures 5(a) and (b), we can observe that cmOCT using the modified scanning protocol provides better signal to noise ratio than SV technique using the same B-M scan protocol. Table 3 provides the CSNR, SNR and CNR measurements for SV and cmOCT techniques. For calculation of CSNR, SNR and CNR, flow region signals were obtained from the profile plot marked in yellow passing through the hair pin loops and noise region signals were obtained from the profile plot marked in white as shown in figure 5(a) and 5(b). For comparison between both techniques, the profile plots were mapped to the range 0 -1. From the profile plots shown in figure

5(c), we can observe that noise is reduced in cmOCT mapping method compared to SV technique. From table 3, we can observe that there is an improvement of four- fold in CSNR calculation and 10 dB in SNR calculation for cmOCT technique compared to SV technique.

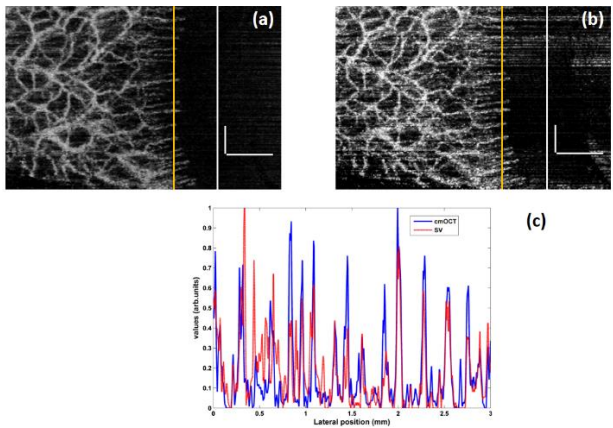

Fig. 5. (a) MIP of enface cmOCT images from nailfold (b) MIP of SV enface images (c) signal profiles extracted from the position marked in yellow in (a) and (b); Scalebars are 500 µm.

Table 3. Comparison metric for motion contrast obtained from nailfold

| Comparison Metric | cmOCT | SV |
| --- | --- | --- |
| CSNR | 2.8 | 0.61 |
| SNR | 67.1 dB | 55.8 dB |
| CNR | 2.33 | 0.91 |

Also, to evaluate the proposed method for ocular studies, angiographic studies on a rodent model were carried out. This is the first study that has tested the suitability of cmOCT technique to ocular studies. Monitoring ocular circulation is vital as many of the vision related issues begin with changes in the microcirculatory flow or the vascular pattern. Since the introduction of OCT, it has found tremendous applications in ophthalmology and has led to commercialization of the technology with FDA approved clinical ophthalmic OCT systems widely in use both in research and in clinics. However, it is to be noted that very few studies have been carried out to investigate the blood perfusion and vasculature of iris tissue beds compared to retinal angiography. Few studies have been conducted that evaluated the suitability of commercial ophthalmic spectral domain OCT systems during disease conditions such as iris microhemangiomatosis, iris nevus, iris hemangioma [41]–[43], normal iris vasculature [44], in staging neovascularization [45] and in iris tumors [46]. Most of these studies are based on spectral domain systems operating at 70 kHz A-scan rate with an inbuilt SSADA technique. Iris vasculature studies have been performed using the OMAG based technique on rodent models using spectral domain OCT system [47]. Similar studies on rodent models have been carried out using photoacoustic microscopic imaging [48], [49].

The current gold standard for ocular angiography is fluorescence angiography. One of the drawbacks of fluorescence angiography is that the imaging requires injection of contrast agents, usually fluorescein or indocyanine green. This causes pain, discomfort and in some cases allergic reactions due to the injected dye. As angiographic OCT provides label free endogenous contrast mechanism from the moving RBC's, it has the potential in replacing fluorescence angiography. Based on these observations, we performed cmOCT angiography on a preclinical rat model. For the experiments, the rat was mounted on an in-house built imaging platform that minimized the head movement and the blinking of the eye. The preclinical rat model was anaesthetized during the entire imaging session using a mixture of 1.5% isoflurane, 80 % air, and 20 % oxygen inhaled through the attached breathing mask. The body temperature of the animal was maintained constant at 37 °C using a temperature-controlled heating pad. For the data acquisition, B-M mode scanning protocol with 8 scans at each location, with 200 transverse scans covering 5 mm x 5 mm was performed. Throughout the experiment, the focal plane of imaging was fixed midway between the iris. The 3D volume data was acquired in 27 s. Figure 6 shows the results of the experiment. Figure 6(a) shows a representative OCT B-scan image formed by averaging the repeated B-scan images at a given position and figure 6(b) shows the corresponding averaged 2-D correlation map as calculated from section 2.2. Figure 6(d) and 6(f) shows the depth encoded intensity projection image and MIP of enface images of the vasculature over the entire scanned area using cmOCT technique. From the B-scan image in figure 6(a), one can see the structures within the anterior segment of the eye namely the avascular cornea and highly vascularized iris. Since the focal plane was fixed midway of the iris, we can observe the mirror artefact that occurs in FD-OCT systems. Figures 6(c) and (e) and (g), shows the corresponding SV image of the B-scan, the depth encoded intensity projection image and MIP of enface images obtained by SV technique respectively.

It can be seen that the cmOCT is capable of being applied to high speed OCT systems with wide field of view imaging applications comparable to OMAG based angiography reported by Choi et al [47] for a smaller field of view. The dense small capillaries of the iris which are typically less than 10 µm could not be well resolved by our low NA objective lens. However, the general micrvasculature of the rat iris consisting of radial branches of vessels and dense capillary network surrounding the pupil is clearly visible in figure 6(d). Also, we can see the major iris circle (MIC), which bifurcates into multiple radial iris arteries (RIA) which then feed into the lesser iris circles (LIC) around the pupillary margin. It is to be noted that some of the RIA and the LIC appear to be slightly blurred. This could be due to the combined effect of low objective NA and degradation of signal quality of the structures out of the focal plane.

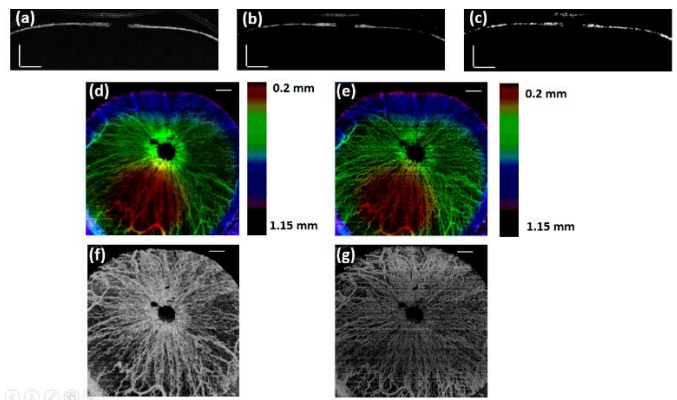

Fig. 6. (a) Representative B scan image of anterior chamber of rat, (b) corresponding cmOCT image, (c) corresponding SV image (d) depth encoded intensity projection image of enface cmOCT maps, (e) depth encoded intensity projection image of enface cmOCT maps, (f) – (g) MIP of enface images of cmOCT and SV (scale bars is 500 µm).

Figures 6 (f) and (g) indicates reduced signal to noise ratio of SV images compared to cmOCT images. However, the structures within the vasculature namely MIC, RIA and LIC are visible with reduced contrast.

Next, we imaged a rodent eye that had undergone a failed corneal grafting procedure. Cornea is the transparent, smooth avascular region within the anterior segment of the eye and plays an important role in focussing light onto the retina. Corneal layer can be damaged or scarred by factors such as chemical injury, infections or genetic dystrophy. Corneal grafting or penetrating keratoplasty is one of the treatment approaches undertaken when the corneas have become opaque. However, it has been reported that, in 35% of cases undergoing the procedure, there is a possibility of graft rejection [50], [51]. Major clinical symptoms of graft rejection include corneal edema, stromal infiltration, and corneal vascularization. OCT offers great potential in the treatment monitoring of such grafted corneas over time. Introducing corneal vascularization can be used as a tool to study the neovascularization process in general and in particular within the eye for clinical researchers. High resolution, high speed OCT's have the potential to be used as a non-invasive imaging tool to study the functional processes over time that lead to neovascularization as well as the structural changes that occur. In view of this, we imaged a rat eye with a failed corneal graft. cmOCT was able to capture the newly formed vessels within the cornea. Figure 7(a) shows a representative B scan image obtained from the failed transplant model and figure 7(b) shows the corresponding cmOCT map. Corneal inflammation and penetration of the iris into the cornea along with loss of corneal transparency can be observed from figure 7(a). From the corresponding cmOCT image shown in figure 7(b), we can see the capillary vessels within the cornea that lead to corneal vascularization. Finally, figure 7(d) shows the depth encoded intensity projection images obtained from the cmOCT *enface* images. The projection map enables the visualization of the vasculature within the cornea that may be of interest to clinicians and researchers studying vascularization/neovascularization of the cornea following corneal injury or treatment models and also in studying the vascular development during ocular tumor or any other cancerous tissue. Figures 7(f) and (g) shows the MIP of enface images from cmOCT and SV techniques respectively.

Fig. 8 shows the profile plots obtained from the region marked in yellow in Fig. 7(b) and 7(c) showing detected capillaries by cmOCT and SV angiographic techniques.

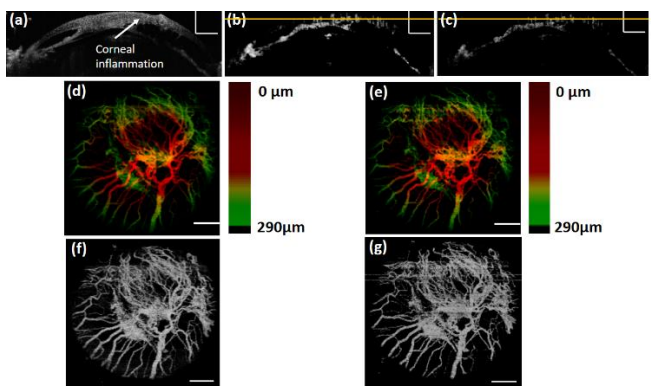

Fig. 7. (a) Representative B scan image of the anterior chamber with a failed corneal graft, (b) corresponding cmOCT image (c) corresponding SV image (d)-(e) depth encoded intensity projection image of enface cmOCT and SV maps (f) –(g) MIP of enface images of cmOCT and SV (scale bars is 500 µm).

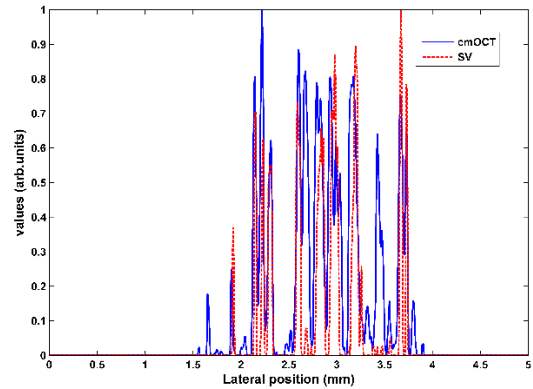

Fig. 8. Signal profile extracted from the position marked in yellow in fig. 7(b) and (c).

From the profile plots, we can observe that, cmOCT detects more capillaries (peak signals) compared to SV.

Theoretically, the resolution in the B-scan plane used for cmOCT kernelling is worse by a factor of five in each dimension for a 5x5 kernel. The lateral resolution in the other dimension is unaffected. However, we noticed that we display the image in the original resolution and notice that it is better than expected, possibly due to oversampling. We also notice that vessels significantly smaller than the theoretical resolution are easily visible. The axial resolution in tissue is better than that in air by a factor equal to the refractive index (approx. 1.4).

From the above experiments and results, we demonstrate the suitability of cmOCT angiography in high speed SS-OCT systems.

## 4 CONCLUSIONS

In this article, we presented and described the applicability of the cmOCT angiography incorporating modified scanning protocol using high-speed SS-OCT. Also, we demonstrated the suitability of the cmOCT technique to *in vivo* ocular imaging in the rodent model for the first time. The modified scanning protocol based on acquiring 8 scans at each location is shown to produce angiographic maps with better SNR compared to speckle variance algorithm that used the same number of multiple B-scans for the calculation of variance. In this paper, we have not fully investigated as to why cmOCT gives better contrast compared to SV imaging. A possible reason could be that correlation between two frames in time undergoing movement provides better contrast compared to taking the variance of the frames in time. Also, here we have produced an average cmOCT map obtained from *n-1* cmOCT maps where *n* is the number of B – scans acquired at the same location over time. However, in SV processing, we have used the entire *n* B – scans to calculate the SV values. However, unlike SV, cmOCT can be operated on just two B-scans, especially if SNR is good, where faster scanning and processing is desired. Further studies have to be carried out to optimize cmOCT technique for high speed SS-OCT by optimizing the B-M scan protocol. The technique presented using 200 kHz SS-OCT system operating at 1300 nm offers great potential in clinical imaging as it can be used to rapidly acquire 3D-OCT volumes over a wide field of view and larger penetration depths without compromising structural-angiographical details.

ACKNOWLEDGEMENT

This work was supported by the National Biophotonics Imaging Platform Ireland funded under the Higher Education Authority PRTLI Cycle 4, cofunded by the Irish Government and the European Union–Investing in your future.